\documentclass[10pt,conference,compsocconf,letterpaper]{IEEEtran} 

\usepackage{multirow}
\usepackage{multicol}
\usepackage{comment}
\usepackage[pdftex]{color, graphicx}
\usepackage{subfig}
\usepackage{amsmath}
\usepackage{amssymb}
\usepackage[vlined]{algorithm2e}
\usepackage{url}
\usepackage{xcolor}
\usepackage{float}
\usepackage[T1]{fontenc}
\usepackage[utf8]{inputenc}
\usepackage{soul}
\usepackage[normalem]{ulem}
\usepackage[space]{cite}

\newcommand{\fixme}[1]{{\color{red}\em\bf{[FIXME: #1]}}}

\begin{document}



\title{A Comparative Study of Pre-trained Speech and Audio Embeddings for Speech Emotion Recognition}


\author{
\IEEEauthorblockN{Orchid Chetia Phukan}
\IEEEauthorblockA{\textit{Dept. of CSE} \\
IIIT Delhi, India\\
orchidp@iiitd.ac.in}
\and
\IEEEauthorblockN{Arun Balaji Buduru}
\IEEEauthorblockA{\textit{Dept. of CSE} \\
IIIT Delhi, India\\
arunb@iiitd.ac.in}
\and
\IEEEauthorblockN{Rajesh Sharma}
\IEEEauthorblockA{\textit{Institute of Computer Science} \\
University of Tartu, Estonia\\
rajesh.sharma@ut.ee}
}

\maketitle

\begin{abstract}

Pre-trained models (PTMs) have shown great promise in the speech and audio domain. Embeddings leveraged from these models serve as inputs for learning algorithms with applications in various downstream tasks. One such crucial task is Speech Emotion Recognition (SER) which has a wide range of applications, including dynamic analysis of customer calls, mental health assessment, and personalized language learning. PTM embeddings have helped advance SER, however, a comprehensive comparison of these PTM embeddings that consider multiple facets such as embedding model architecture, data used for pre-training, and the pre-training procedure being followed is missing. A thorough comparison of PTM embeddings will aid in the faster and more efficient development of models and enable their deployment in real-world scenarios. In this work, we exploit this research gap and perform a comparative analysis of embeddings extracted from eight speech and audio PTMs (wav2vec 2.0, data2vec, wavLM, UniSpeech-SAT, wav2clip, YAMNet, x-vector, ECAPA). We perform an extensive empirical analysis with four speech emotion datasets (CREMA-D, TESS, SAVEE, Emo-DB) by training three algorithms (XGBoost, Random Forest, FCN) on the derived embeddings. The results of our study indicate that the best performance is achieved by algorithms trained on embeddings derived from PTMs trained for speaker recognition followed by wav2clip and UniSpeech-SAT. This can relay that the top performance by embeddings from speaker recognition PTMs is most likely due to the model taking up information about numerous speech features such as tone, accent, pitch, and so on during its speaker recognition training. Insights from this work will assist future studies in their selection of embeddings for applications related to SER.


\end{abstract}

\section{Introduction}
\label{intro}
Pre-trained models (PTMs) are widely available in the speech and audio signal processing domain. Pre-training is carried out on large-scale speech (Librispeech (LS) \cite{panayotov2015librispeech}) or non-speech (AudioSet (AS) \cite{gemmeke2017audio}, VGGSound (VS) \cite{chen2020vggsound}) databases. They find application in various narrow-domain tasks in different ways: from feature extractors for downstream models to the whole model being fine-tuned on task-specific data. Their model architectures can be of varied nature, it can be Convolution Neural Network (CNN) based such as AlexNet, VGG, Inception, ResNet \cite{hershey2017cnn}, etc., and also, attention-based such as AALBERT \cite{chi2021audio}, CAV-MAE \cite{gong2022contrastive}, etc. Pre-training is executed using different approaches: supervised \cite{DBLP:journals/corr/abs-2104-01778} or self-supervised fashion \cite{liu2021tera}. Embeddings exploited from PTMs are used for different tasks, for example, covid-19 detection \cite{campana20223}, music emotion recognition \cite{koh2021comparison}, speech emotion recognition (SER) \cite{macary2021use}. \par

In this work, we focus on SER, an important task for human-machine interaction. It has gained traction in recent times due to its prospective applications in a wide span of different domains, for instance, psychology, healthcare, and fields that often include customer interactions, such as customer service providers, call centers, and so on. A variety of methods have been applied for SER, ranging from fuzzy methods \cite{razak2005comparison}, Hidden Markov Model (HMM) based methods \cite{vlasenko2007tuning}, classical machine learning-based approaches \cite{iliou2009comparison}, deep learning-based methods\cite{trigeorgis2016adieu} to embeddings from PTMs such as wav2vec 2.0 \cite{pepino2021emotion}, HuBERT \cite{pastor2022cross}. The availability of a large number of PTMs has resulted in significant progress in the field of SER. As they were trained on vast amounts of data and learned detailed and nuanced representations of speech, the embeddings extracted from them have proven beneficial for emotion recognition. 

However, it is not clear which PTM embeddings are best for SER. Keesing et al. \cite{keesing2021acoustic} provided a comparison between acoustic and neural (speech and audio) embeddings by training downstream classifiers such as SVM, RF, MLP, etc. on various speech emotion databases. Atmaja et al. \cite{atmaja2022evaluating} assessed representations of PTMs for SER that were pre-trained in a self-supervised manner on speech data by training an FCN classifier on the representations as input features. But a comprehensive comparison of embeddings extracted from a broad variety of PTMs with consideration of their model architectures, pre-training methodologies, and pre-training datasets has not been carried out for SER. We address this research gap by conducting a comparative study of embeddings extracted from eight PTMs by training low-level models with the embeddings as input features. \par

To summarize, the following are our main contributions: 

\begin{itemize}

  \item Compiling PTM embeddings that could be useful for performing downstream SER tasks. We consider many diverse PTMs (wav2vec 2.0, data2vec, wavLM, UniSpeech-SAT, wav2clip, YAMNet, x-vector, ECAPA) with varied model architectures, pre-training data, and pre-training procedures.
  \item Comprehensive comparative analysis of different PTM embeddings through downstream classifiers (XGBoost, Random Forest, Fully Connected Network) which are trained and evaluated on four public datasets (CREMA-D, TESS, SAVEE, Emo-DB).
  \item Our study has found that embeddings from PTMs trained for speaker recognition tasks perform better than embeddings from other categories of Speech/Audio PTMs. Our hypothesis is that this could be speaker recognition training procedures enabling models to learn various aspects of speech such as tone, accent, pitch.
  
\end{itemize} 

\noindent This paper is divided into six sections. Section \ref{back} discusses past works on PTMs followed by Section \ref{data} which elaborates on the different speech emotion databases taken into consideration for carrying out our experiments. In Section \ref{aptm}, we provide brief information on PTM embeddings considered for our analysis and the reason behind the consideration. Section \ref{exp} focuses on the lower-level classifiers, their implementation, training, and results obtained for the comparative analysis. Finally, Section \ref{conc} concludes the work presented and gives prospective directions for future work.

\section{Related Works}

\label{back}

\begin{table*}[hbt!]
\centering
\caption{Basic information related to various speech emotion corpora}
\label{tab:basicinfo}
\begin{tabular}{lllll}   
\hline
\textbf{Corpus} & \textbf{Lanaguage} & \textbf{\# of utterances} & \textbf{\# of speakers} & \textbf{Labeled Emotions}\\\hline
CREMA-D & English & 7442 & 91 & Anger, Happiness, Sadness, Fear, Disgust, Neutral \\ 
TESS & English & 2800 & 2 &  Anger, Happiness, Sadness, Fear, Disgust, Neutral, Surprise \\ 
SAVEE & English & 480 & 4 & Anger, Happiness, Sadness, Fear, Disgust, Neutral, Surprise  \\ 
Emo-DB & German & 535 & 10 & Anger, Happiness, Sadness, Fear, Disgust, Neutral, Bored  \\ 
\hline
\end{tabular}
\end{table*}
Initially, PTM architectures were mostly CNN-based, for instance, SoundNet \cite{aytar2016soundnet}, a 1D CNN trained on a massive amount of unlabeled videos collected from Flickr. It was trained in collaboration with a visual recognition network via discriminative knowledge transfer. Later, the trained model's representations were used as features combined with posterior classifiers to classify acoustic scenes. With the availability of a large-scale labeled audio dataset, AS, various models such as VGGish \cite{hershey2017cnn}, L3-Net \cite{cramer2019look}, PANNs \cite{kong2020panns}, and etc. were proposed. VGGish is based on VGG architecture and was trained in a supervised manner to classify 527 sound events. L3-Net is also based on the VGG network and was pre-trained in a self-supervised manner for audio-visual correspondence. 
Gong et al. \cite{gong2021psla} trained EfficientNet for audio tagging on AS that was first trained on ImageNet (IM). They also discussed how pre-training in a different modality boosts performance. Niizumi et al. \cite{https://doi.org/10.48550/arxiv.2103.06695} extended Bootstrap your own latent (BYOL) approach initially given for vision to BYOL for audio (BYOL-A). BYOL-A presents a novel generalized self-supervised approach for generating audio representation and employs a CNN as an encoder. It was pre-trained on AS by removing the labels and achieved competitive results on various low-level tasks such as speaker identification, language identification, etc. with baseline models. Schneider et al. \cite{DBLP:journals/corr/abs-1904-05862} proposed a novel pre-trained multilayer CNN model wav2vec, trained on unlabeled speech data for speech recognition. wav2vec reported the lowest WER for character-based speech recognition compared to past works. \par

Mockingjay \cite{Liu_2020}, a multi-layer bidirectional transformer model was pre-trained on LS using masked modeling, where 15\% of the input frames were masked to zero and it outputs the masked frames. They observed that pre-training Mockingjay in this manner resulted in improved performance in downstream supervised activities. Baevski et al. proposed wav2vec 2.0 \cite{https://doi.org/10.48550/arxiv.2006.11477}, where the initial layer is a convolutional layer that acts as a feature encoder followed by transformer layer. It is trained in a self-supervised way where masking of a few parts of the feature encoder outputs is done. Unlabeled LS is used as pre-training data and it improves upon wav2vec for phoneme recognition. HuBERT \cite{DBLP:journals/corr/abs-2106-07447}, a BERT-like architecture with self-supervised training was also devised that achieves comparable performance with wav2vec 2.0 for speech recognition in LS. The first fully attention-based convolution-devoid architecture named Audio-Spectrogram transformer (AST) was presented in \cite{DBLP:journals/corr/abs-2104-01778} for audio classification tasks. It accepts mel-spectrogram as input. AST uses the advantages of pre-trained ViT for image classification tasks, and it is afterward trained on AS. Over previous investigations, AST reported state-of-the-art (SOTA) performance on the AS, ESC-50, and Speech Commands V2 databases. Gong et al. \cite{DBLP:journals/corr/abs-2110-09784} enhanced AST further by training it in a self-supervised pattern through joint discriminative training and masked spectrogram modeling. This kind of training improved performance in lower-level tasks over the supervised version. 
Various encoder-decoder architectures, such as Audio-MAE \cite{https://doi.org/10.48550/arxiv.2207.06405} and MaskSpec \cite{https://doi.org/10.48550/arxiv.2204.12768}, was also proposed. \par 

Embeddings from PTMs such as YAMNet and wav2vec trained on audio and speech data, respectively, were used as input features to classifiers for SER \cite{keesing21_interspeech}. Using models pre-trained on large databases and exploiting embeddings of them as features and applications of transfer learning by finetuning holds a promising future for SER. However, no comparison of embeddings recovered from a wide range of PTMs has been conducted for SER taking into account their model architectures, pre-training procedures, and pre-training datasets. To fill this knowledge gap, we conduct a comparative investigation of embeddings retrieved from eight diverse PTMs pre-trained on speech and audio data. 

\begin{figure}[hbt!]
\centering
    \subfloat[CREMA-D]{{\includegraphics[height = 0.33\linewidth, width=0.4\linewidth]{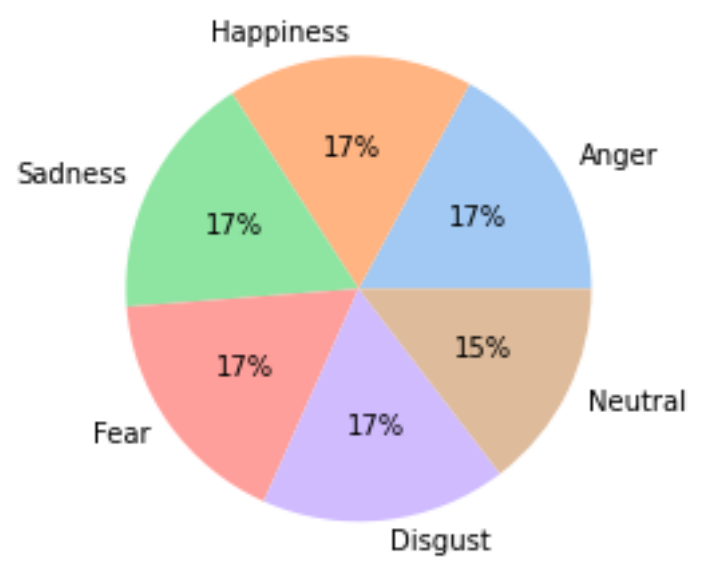}}\label{fig:cremademo}}\quad
    \subfloat[TESS]{{\includegraphics[height = 0.33\linewidth, width=0.4\linewidth]{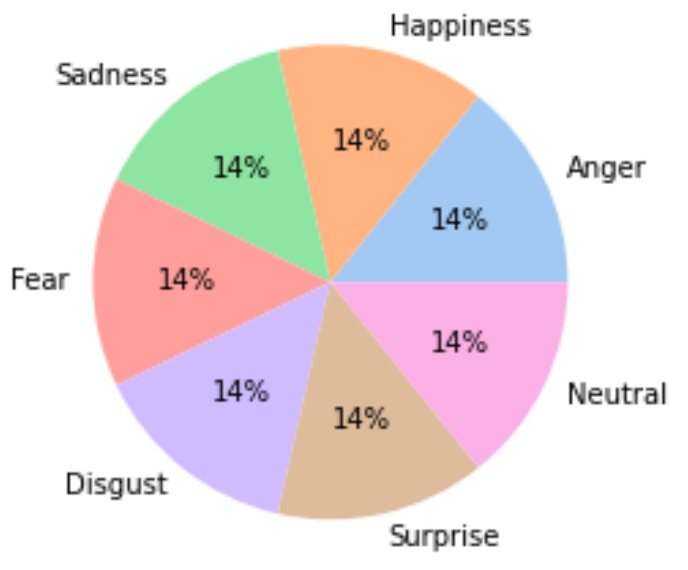}}\label{fig:tessemo}}\quad
    \subfloat[SAVEE]{{\includegraphics[height = 0.33\linewidth, width=0.42\linewidth]{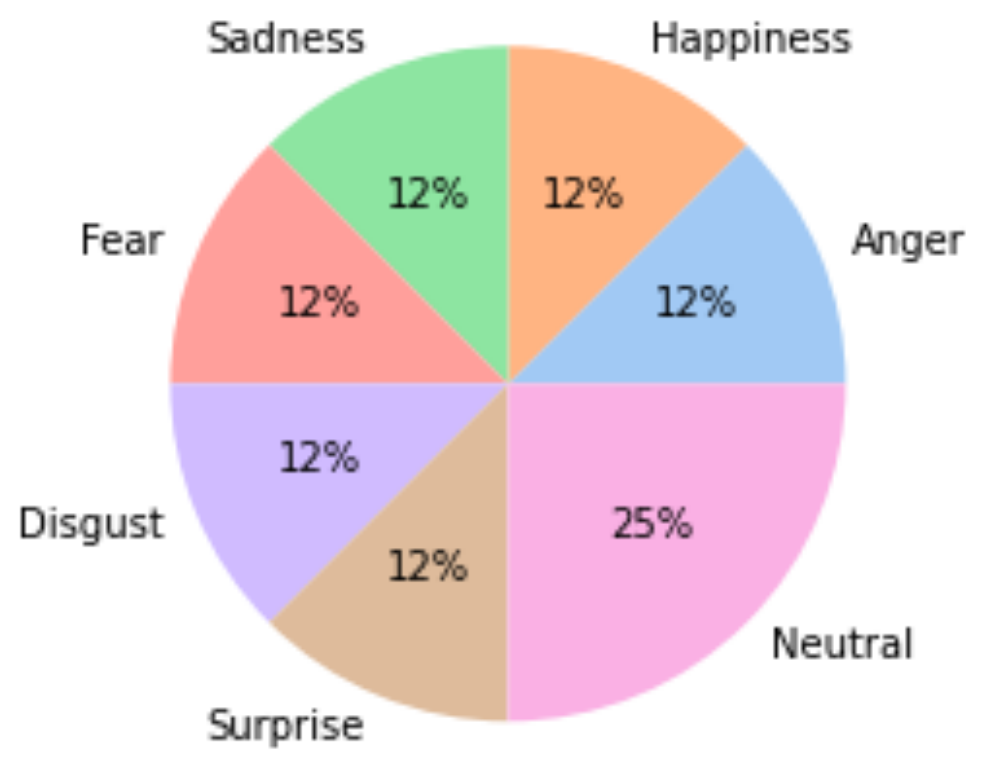}}\label{fig:saveeemo}}\quad
    \subfloat[Emo-DB]{{\includegraphics[height = 0.33\linewidth ,width=0.4\linewidth]{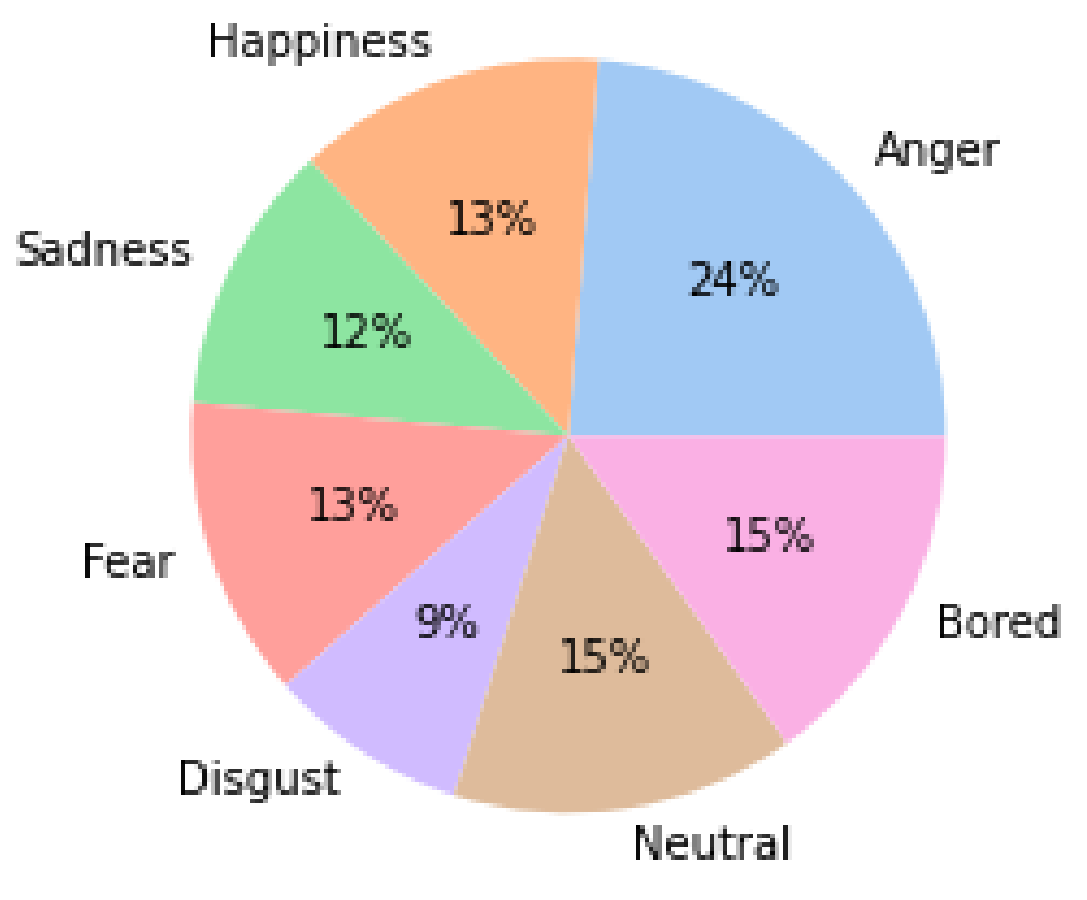}}\label{fig:emodbemo}}
   \caption{Distribution of Emotions across different corpora}
\label{fig:cemo}
\end{figure}

\section{Speech Emotion Corpora}

\label{data}

We experiment with four openly accessible benchmark speech emotion databases: Crowd-Sourced Emotional Multimodal Actors Dataset (CREMA-D) \cite{cao2014crema},  Ryerson Audio-Visual Database of Emotional Speech and Song (RAVDESS) \cite{livingstone2018ryerson}, Toronto Emotional Speech Set (TESS) \cite{utorontoTorontoEmotional}, Surrey Audio-Visual Expressed Emotion (SAVEE) \cite{jackson2014surrey}, and German Emotional Speech Database (Emo-DB) \cite{burkhardt2005database}. Essential information and distribution of emotions for each corpus are given in Table \ref{tab:basicinfo} and Figure \ref{fig:cemo} respectviely. \par 

Additional information related to the databases can be found below:
\begin{itemize}

  \item\textbf{CREMA-D:} The audio snippets feature 48 male and 43 female performers from various ethnic origins. They talked from a list of 12 phrases.  With a diverse range of ages, genders, and ethnicities, CREMA-D is a high-quality data source for SER.
  \item\textbf{TESS:} It is recorded by two female actors. Both actresses were fluent in English and cherry-picked 200 words were spoken by the actresses for the seven emotions.
  \item\textbf{SAVEE:} It comprises recordings of four male actors in British English accents. For each emotion, the actors delivered phrases that were phonetically balanced.
  \item\textbf{Emo-DB:} Recordings are from 5 male and 5 female actors. The actors were given a selection of ten distinct scripts from which to speak.
\end{itemize} 

\section{Pre-trained Model Embeddings}\label{aptm}

\begin{table*}[h!]
\centering
\caption{Comparison of XGBoost trained on different PTM embeddings}
\label{tab:compxgb}
  \begin{tabular}{|l|l|l|l|l|l|l|l|l|}
    \hline
    \multirow{2}{*}{Audio PTM} &
      \multicolumn{2}{c|}{CREMA-D} &
      \multicolumn{2}{c|}{TESS} &
      \multicolumn{2}{c|}{SAVEE}&
      \multicolumn{2}{c|}{Emo-DB}\\
     & Accuracy & F1-score & Accuracy & F1-score & Accuracy & F1-score & Accuracy & F1-score\\\hline
    wav2vec 2.0 & 40.29 & 40.47 & 69.76 & 69.61 & 40.28 & 29.44 & 49.38 & 46.90\\\hline
    data2vec & 49.33 & 49.52 & 76.90 & 76.37 & 37.50 & 29.29 & 49.38 & 48.22\\\hline
    wavLM & 45.48 & 45.85 & 83.10 & 82.41 & 50.00 & 42.73 & 54.32 & 52.06\\\hline
    UniSpeech-SAT & 56.13 & 56.35 & 83.57 & 83.40 & 45.83 & 32.86 & 69.70 & 61.42\\\hline
    wav2clip & 47.45 & 46.77 & 95.00 & 94.95 & 55.56 & 52.79 & 72.84 & 66.31 \\\hline
YAMNet & 46.82 & 46.49 & 92.38 & 92.35 & 50.00 & 41.17 & 58.02 & 51.41\\\hline
x-vector & \textbf{60.16} & \textbf{60.09} & \textbf{97.86} & \textbf{97.77} & \textbf{68.06} & \textbf{62.17} & \textbf{83.95} & \textbf{80.07} \\\hline
ECAPA & 54.34 & 54.02 & 97.14 & 97.05 & 55.56 & 50.09 & 75.31 & 69.70\\\hline
  \end{tabular}
\end{table*}

\begin{table*}[h!]
\centering
\caption{Comparison of Random Forest trained on different PTM embeddings}
\label{tab:comprf}
  \begin{tabular}{|l|l|l|l|l|l|l|l|l|}
    \hline
    \multirow{2}{*}{Audio PTM} &
      \multicolumn{2}{c|}{CREMA-D} &
      \multicolumn{2}{c|}{TESS} &
      \multicolumn{2}{c|}{SAVEE}&
      \multicolumn{2}{c|}{Emo-DB}\\
     & Accuracy & F1-score & Accuracy & F1-score & Accuracy & F1-score & Accuracy & F1-score \\\hline
wav2vec 2.0 & 37.69 & 37.47 & 57.38 & 56.68 & 38.89 & 25.50 & 56.79 & 51.11\\\hline
data2vec & 44.58 & 44.37 & 68.33 & 67.46 & 36.11 & 23.45 & 58.02 & 54.28\\\hline
wavLM & 40.64 & 41.01 & 76.67 & 75.79 & 45.83 & 35.78 & 50.62 & 47.99\\\hline
UniSpeech-SAT & 49.06 & 48.93 & 78.33 & 77.99 & 45.83 & 32.35 & 60.49 & 49.07\\\hline
wav2clip & 44.94 & 44.16 & 94.52 & 94.50 & 59.72 & 55.24 & 67.90 & 63.55\\ \hline
YAMNet & 43.87 & 42.49 & 88.57 & 88.54 & 51.39 & 39.16 & 53.09 & 50.12\\\hline
x-vector & \textbf{52.01} & \textbf{51.64} & 98.33 & 98.28 & \textbf{61.11} & \textbf{49.89} & 81.48 & 78.40 \\\hline
ECAPA & 44.05 & 43.05 & \textbf{98.57} & \textbf{98.45} & 48.61 & 36.09 & \textbf{83.95} & \textbf{80.98}\\\hline
  \end{tabular}
\end{table*}

\begin{table*}
\centering
\caption{Comparison of Fully Connected Network trained on different PTM embeddings}
\label{tab:compfcn}
  \begin{tabular}{|l|l|l|l|l|l|l|l|l|}
    \hline
    \multirow{2}{*}{Audio PTM} &
      \multicolumn{2}{c|}{CREMA-D} &
      \multicolumn{2}{c|}{TESS} &
      \multicolumn{2}{c|}{SAVEE}&
      \multicolumn{2}{c|}{Emo-DB}\\
     & Accuracy & F1-Score & Accuracy & F1-Score & Accuracy & F1-Score & Accuracy & F1-Score \\\hline
wav2vec 2.0 & 46.02 & 45.81 & 84.76 & 84.40 & 41.67 & 31.98 & 60.49 & 57.70\\\hline
data2vec & 53.89 & 53.76  & 86.67 & 86.08 & 43.06 & 33.41 & 64.20 & 63.35\\\hline
wavLM & 55.77 & 55.57 & 95.00 & 94.80 & 50.00 & 32.27 & 62.96 & 59.63\\\hline
UniSpeech-SAT & 64.28 & 64.43 & 96.67 & 96.65 & 61.11 & 49.71 & 82.72 & 79.04\\\hline
wav2clip & 47.18 & 46.92 & 96.90 & 96.79 & 61.11 & 51.81 & 74.07 & 75.42 \\\hline
YAMNet & 48.25 & 48.22 & 96.19 & 96.09 & 55.56 & 41.52 & 61.73 & 59.46\\\hline
x-vector & \textbf{65.80} & \textbf{65.64} & 98.81 & 98.79 & \textbf{70.83} & \textbf{64.90} & 87.65 & 87.01\\\hline
ECAPA & 61.15 & 60.95 & \textbf{99.52} & \textbf{99.50} & 61.11 & 54.11 & \textbf{88.89} & \textbf{87.09}\\\hline
  \end{tabular}
\end{table*}

\begin{figure*}[t!]
    \centering
    \subfloat[CREMA-D]{{\includegraphics[width=0.25\linewidth, height =3.2cm ]{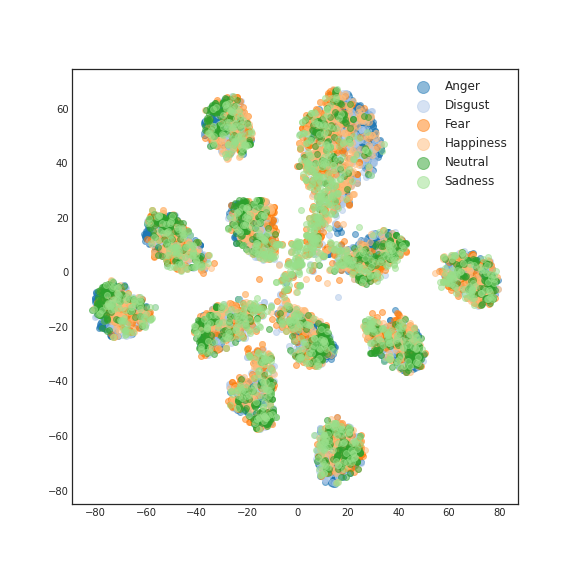}}\label{fig:wav2vec2c}}
    \subfloat[TESS]
    {{\includegraphics[width=0.25\linewidth,height =3.2cm ]{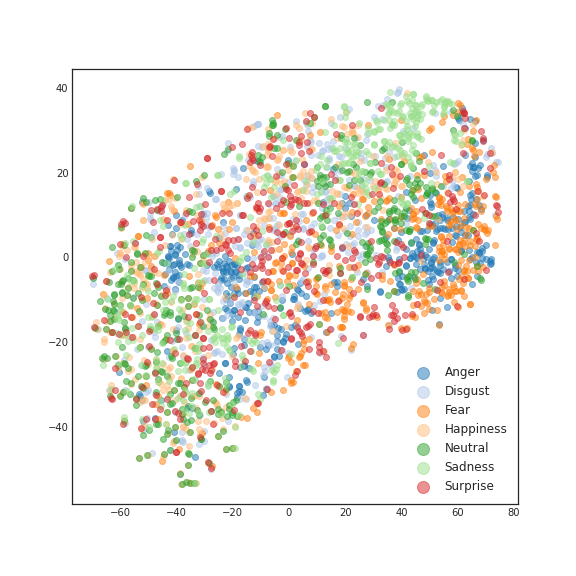}}\label{fig:wav2vec2t}}
    \subfloat[SAVEE]
    {{\includegraphics[width=0.25\linewidth,height =3.2cm ]{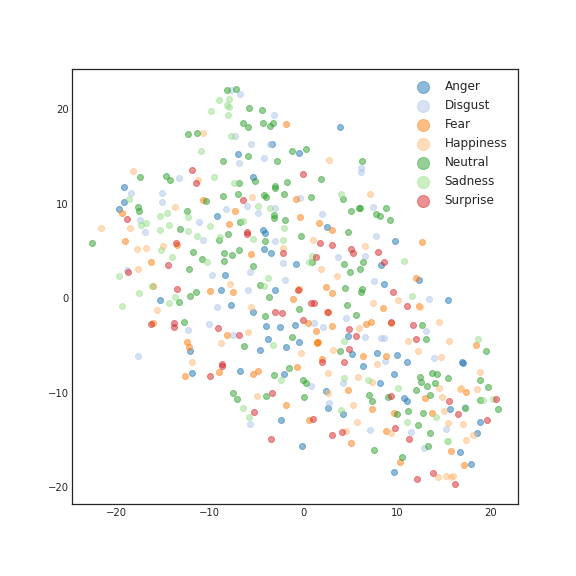}}\label{fig:wav2vec2s}}
    \subfloat[Emo-DB]
    {{\includegraphics[width=0.25\linewidth,height =3.2cm ]{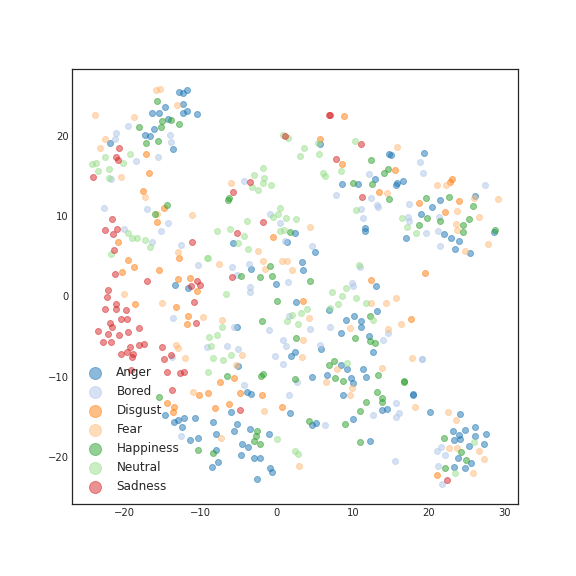}}\label{fig:wav2vec2e}}
   \caption{t-SNE plots of wav2vec 2.0 embeddings across different speech emotion corpora}
\label{fig:tsnewav2vec2}
\end{figure*}

\begin{figure*}[t!]
    \centering
    \subfloat[CREMA-D]{{\includegraphics[width=0.25\linewidth, height =3.2cm ]{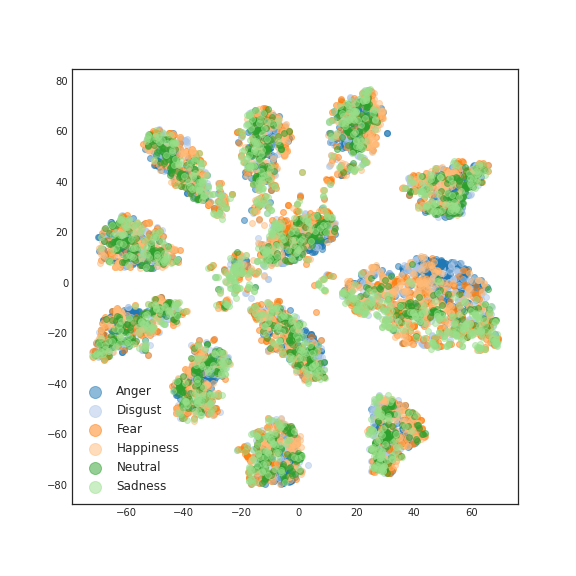}}\label{fig:data2vecc}}
    \subfloat[TESS]
    {{\includegraphics[width=0.25\linewidth,height =3.2cm ]{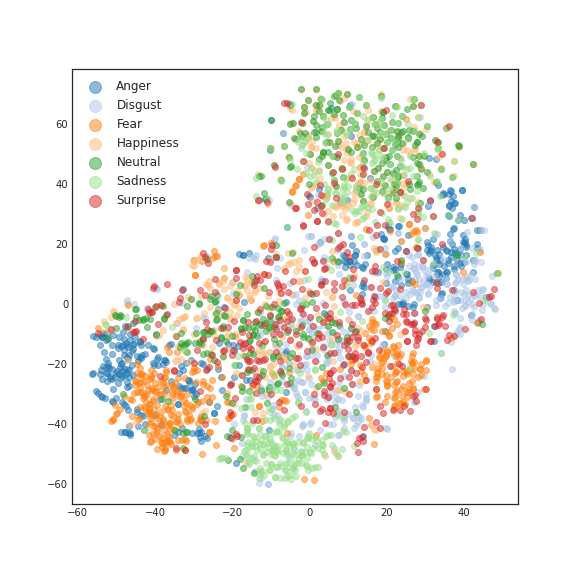}}\label{fig:data2vect}}
    \subfloat[SAVEE]
    {{\includegraphics[width=0.25\linewidth,height =3.2cm ]{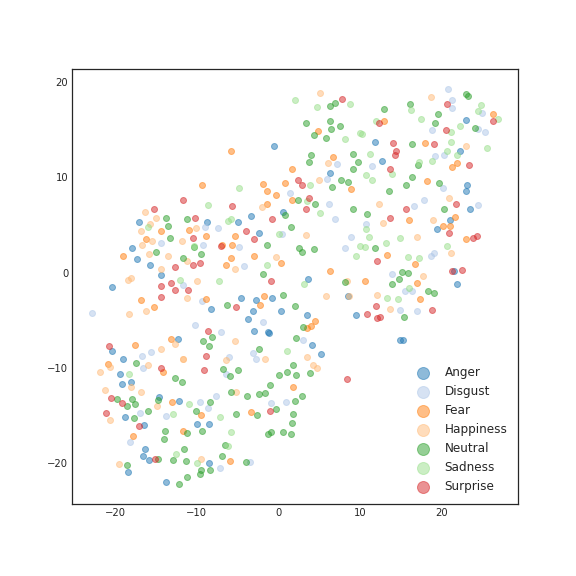}}\label{fig:data2vecs}}
    \subfloat[Emo-DB]
    {{\includegraphics[width=0.25\linewidth,height =3.2cm ]{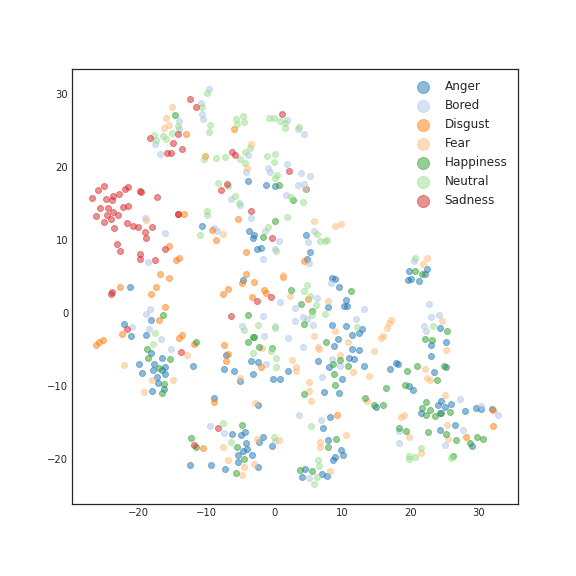}}\label{fig:data2vece}}
   \caption{t-SNE plots of data2vec embeddings across different speech emotion corpora}
\label{fig:tsnedata2vec}
\end{figure*}

\begin{figure*}[t!]
    \centering
    \subfloat[CREMA-D]{{\includegraphics[width=0.25\linewidth, height =3.2cm ]{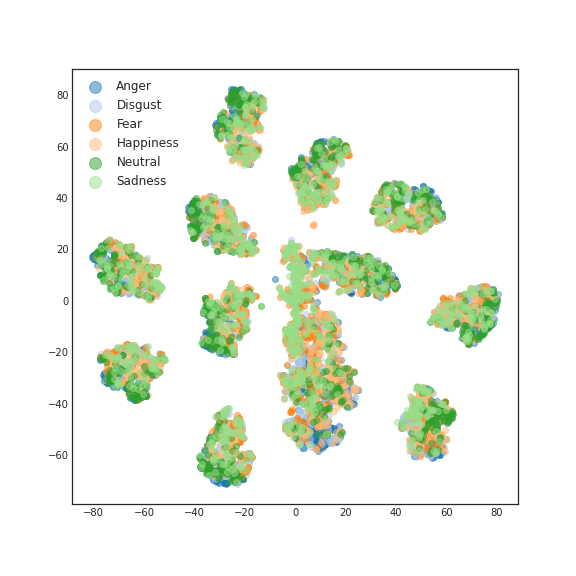}}\label{fig:wavLMc}}
    \subfloat[TESS]
    {{\includegraphics[width=0.25\linewidth,height =3.2cm ]{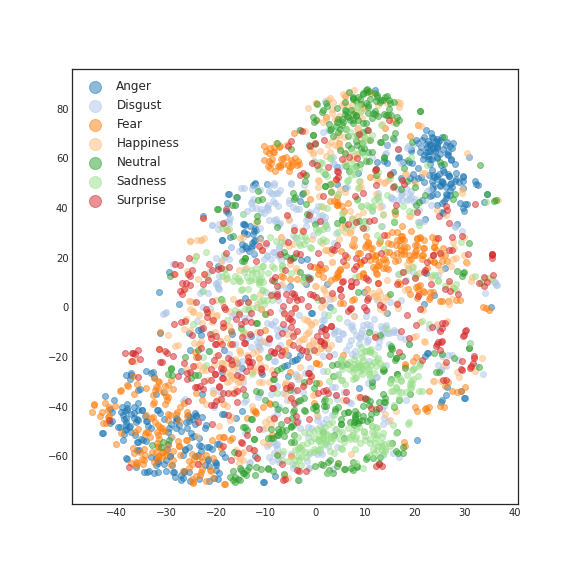}}\label{fig:wavLMt}}
    \subfloat[SAVEE]
    {{\includegraphics[width=0.25\linewidth,height =3.2cm ]{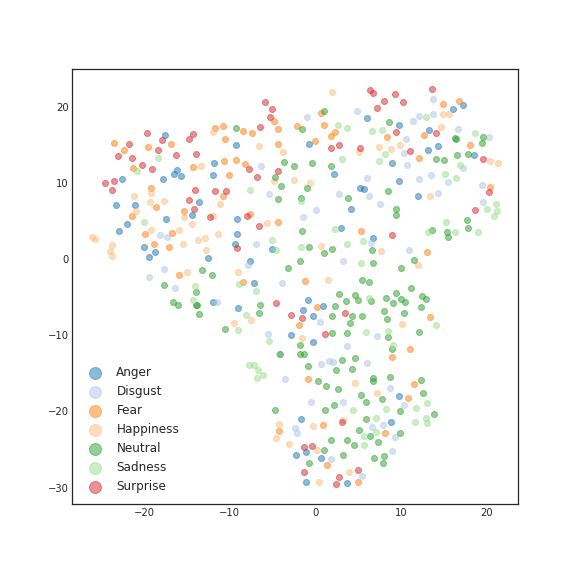}}\label{fig:wavLMs}}
    \subfloat[Emo-DB]
    {{\includegraphics[width=0.25\linewidth,height =3.2cm ]{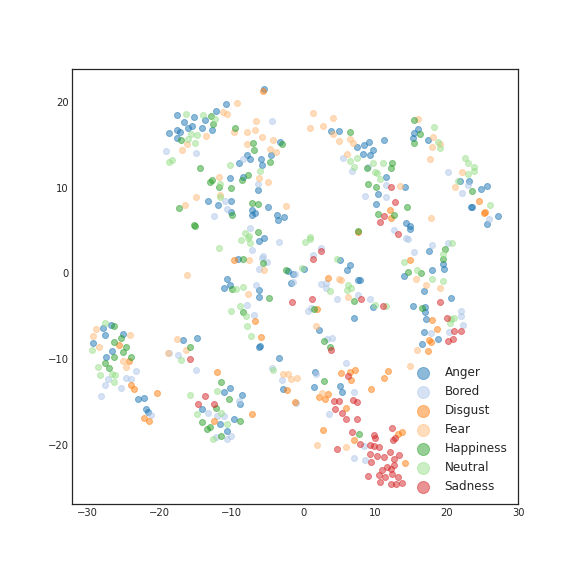}}\label{fig:wavLMe}}
   \caption{t-SNE plots of wavLM embeddings across different speech emotion corpora}
\label{fig:tsnewavLM}
\end{figure*}

\begin{figure*}[t!]
    \centering
    \subfloat[CREMA-D]{{\includegraphics[width=0.25\linewidth, height =3.2cm ]{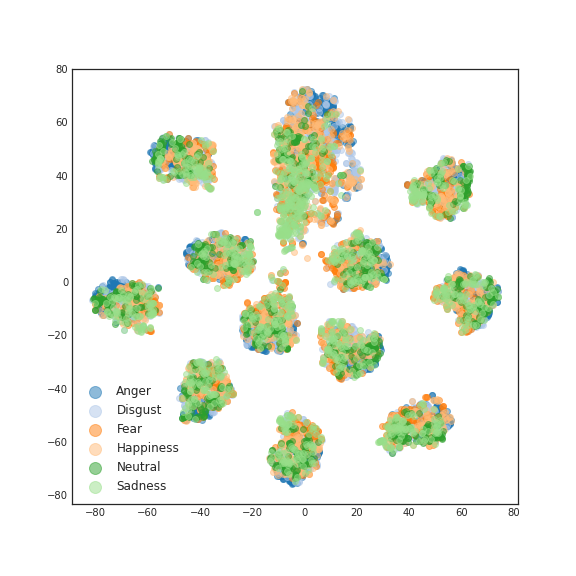}}\label{fig:uc}}
    \subfloat[TESS]
    {{\includegraphics[width=0.25\linewidth,height =3.2cm ]{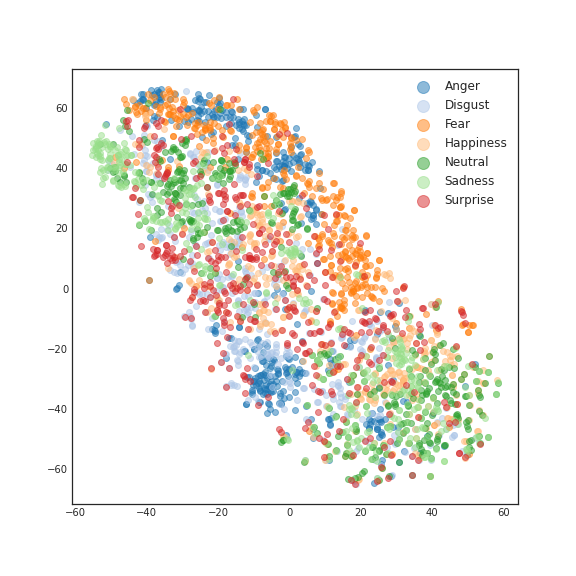}}\label{fig:ut}}
    \subfloat[SAVEE]
    {{\includegraphics[width=0.25\linewidth,height =3.2cm ]{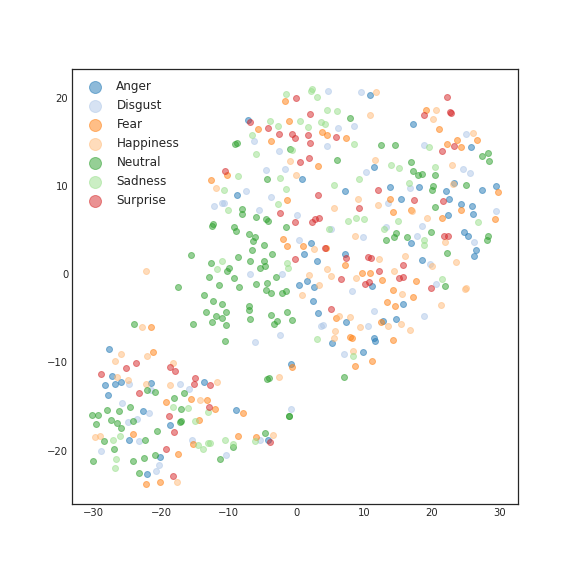}}\label{fig:us}}
    \subfloat[Emo-DB]
    {{\includegraphics[width=0.25\linewidth,height =3.2cm ]{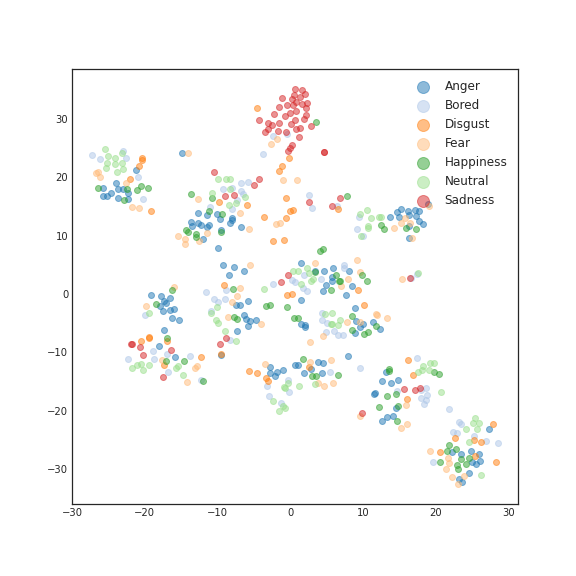}}\label{fig:ue}}
   \caption{t-SNE plots of UniSpeech-SAT embeddings across different speech emotion corpora}
\label{fig:tsneunispeechsat}
\end{figure*}

\begin{figure*}[h!]
    \centering
    \subfloat[CREMA-D]{{\includegraphics[width=0.25\linewidth, height =3.2cm ]{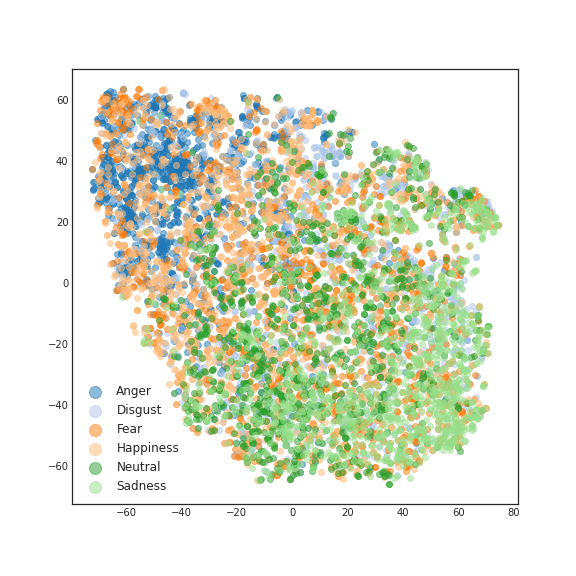}}\label{fig:wav2clipc}}
    \subfloat[TESS]
    {{\includegraphics[width=0.25\linewidth,height =3.2cm ]{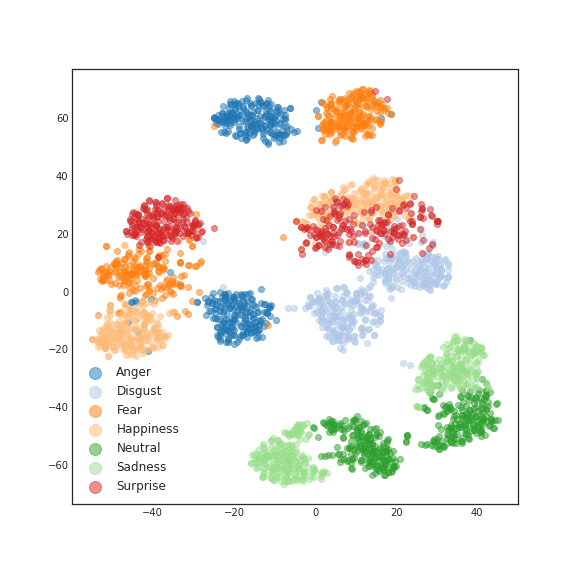}}\label{fig:wav2clipt}}
    \subfloat[SAVEE]
    {{\includegraphics[width=0.25\linewidth,height =3.2cm ]{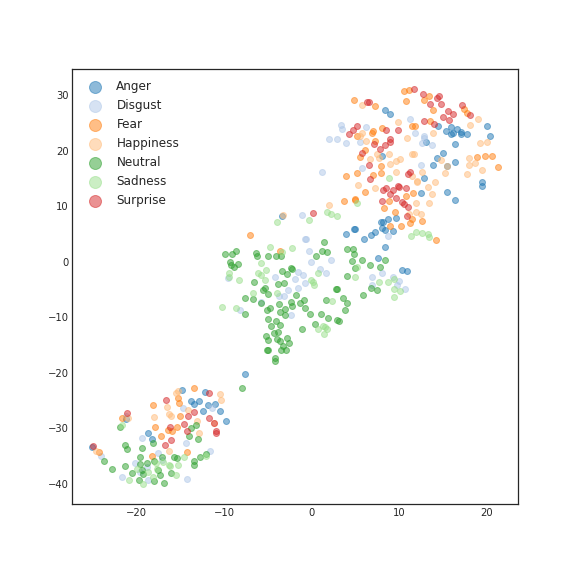}}\label{fig:wav2clips}}
    \subfloat[Emo-DB]
    {{\includegraphics[width=0.25\linewidth,height =3.2cm ]{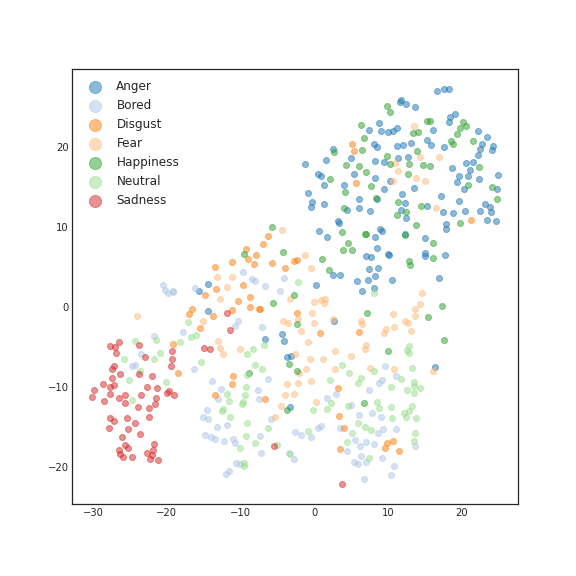}}\label{fig:wav2clipe}}
   \caption{t-SNE plots of wav2clip embeddings across different speech emotion corpora}
\label{fig:tsnewav2clip}
\end{figure*}

\begin{figure*}[t!]
    \centering
    \subfloat[CREMA-D]{{\includegraphics[width=0.25\linewidth, height =3.2cm ]{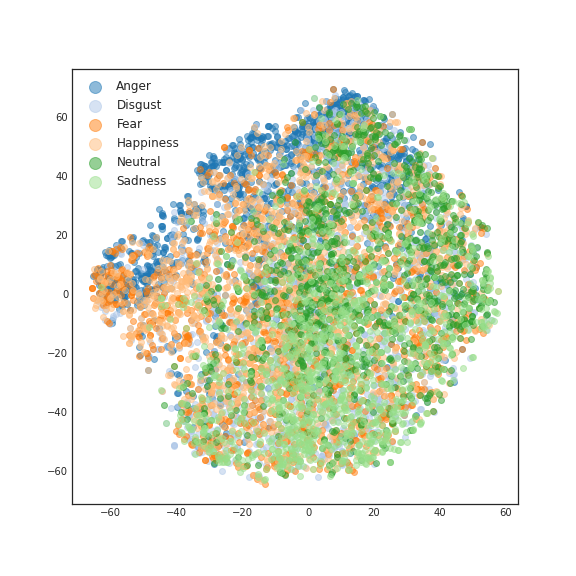}}\label{fig:yamnetc}}
    \subfloat[TESS]
    {{\includegraphics[width=0.25\linewidth,height =3.2cm ]{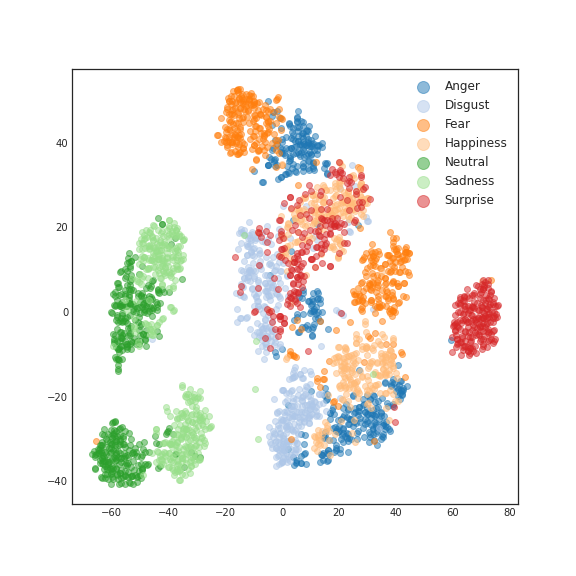}}\label{fig:yamnett}}
    \subfloat[SAVEE]
    {{\includegraphics[width=0.25\linewidth,height =3.2cm ]{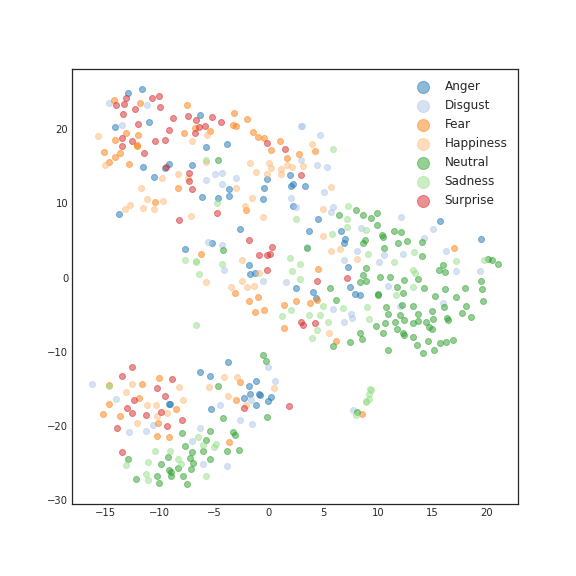}}\label{fig:yamnets}}
    \subfloat[Emo-DB]
    {{\includegraphics[width=0.25\linewidth,height =3.2cm ]{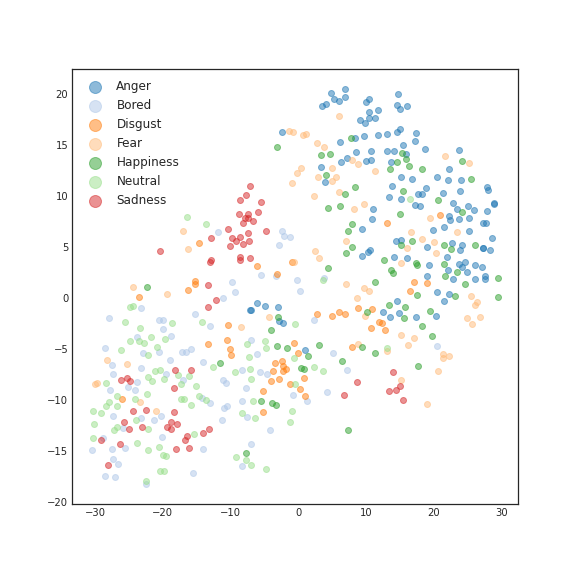}}\label{fig:yamnete}}
   \caption{t-SNE plots of YAMNet embeddings across different speech emotion corpora}
\label{fig:tsneyamnet}
\end{figure*}

\begin{figure*}[t!]
    \centering
    \subfloat[CREMA-D]{{\includegraphics[width=0.25\linewidth, height =3.2cm ]{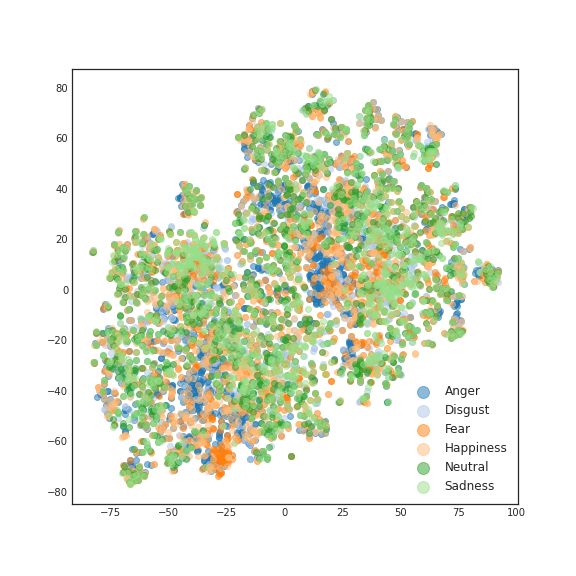}}\label{fig:xvectorc}}
    \subfloat[TESS]
    {{\includegraphics[width=0.25\linewidth,height =3.2cm ]{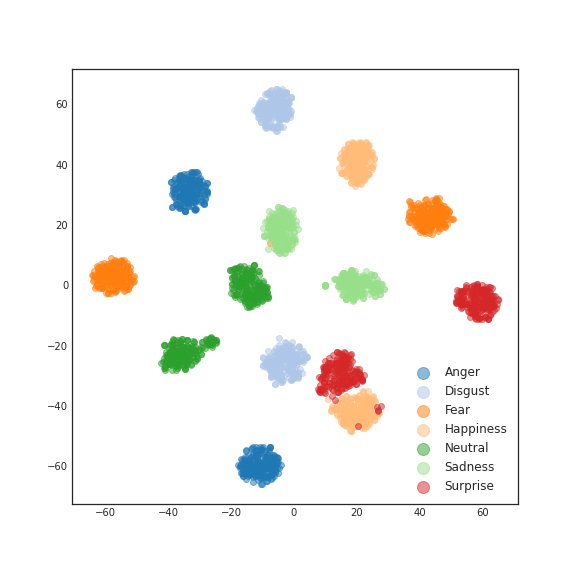}}\label{fig:xvectort}}
    \subfloat[SAVEE]
    {{\includegraphics[width=0.25\linewidth,height =3.2cm ]{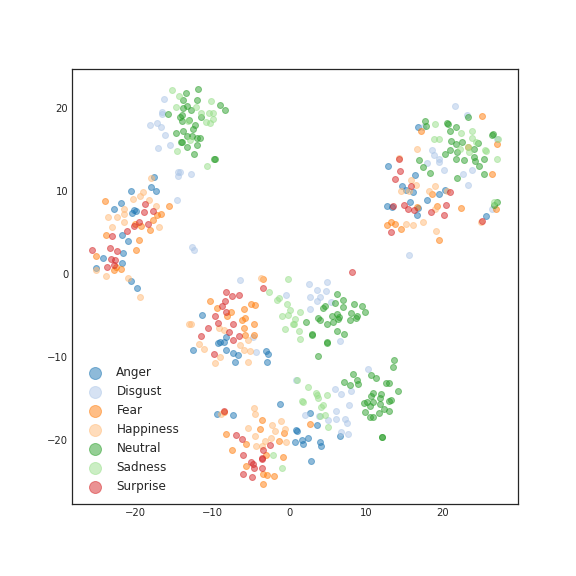}}\label{fig:xvectors}}
    \subfloat[Emo-DB]
    {{\includegraphics[width=0.25\linewidth,height =3.2cm ]{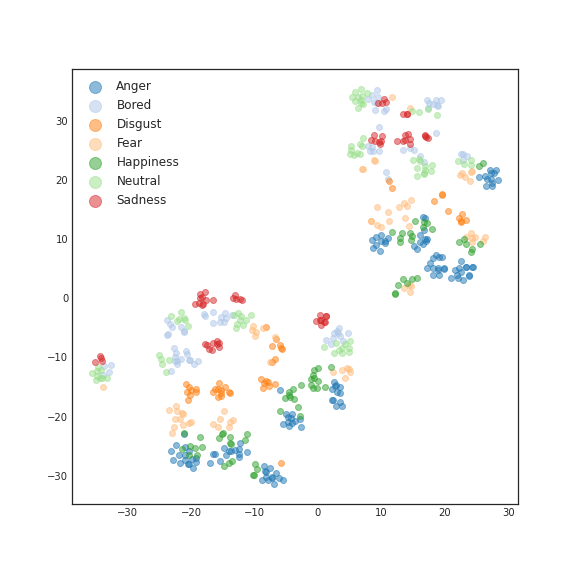}}\label{fig:xvectore}}
   \caption{t-SNE plots of x-vector embeddings across different speech emotion corpora}
\label{fig:tsnexvector}
\end{figure*}

\begin{figure*}[t!]
    \centering
    \subfloat[CREMA-D]{{\includegraphics[width=0.25\linewidth, height =3.2cm ]{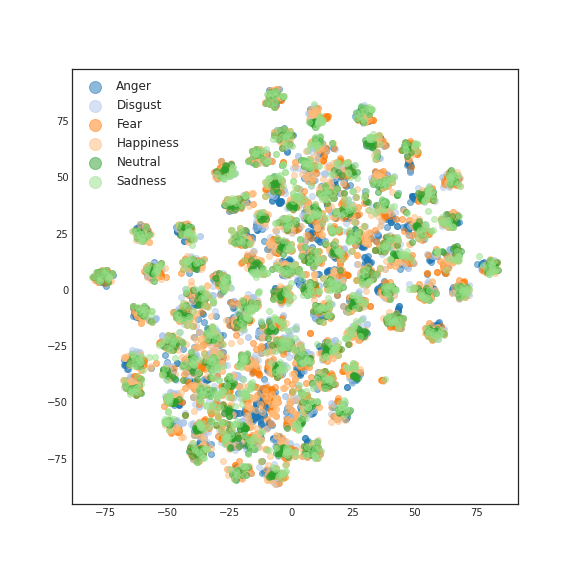}}\label{fig:ecapac}}
    \subfloat[TESS]
    {{\includegraphics[width=0.25\linewidth,height =3.2cm ]{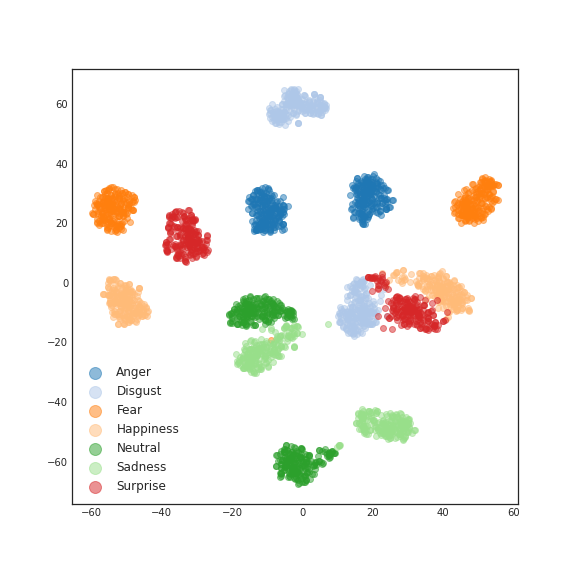}}\label{fig:ecapat}}
    \subfloat[SAVEE]
    {{\includegraphics[width=0.25\linewidth,height =3.2cm ]{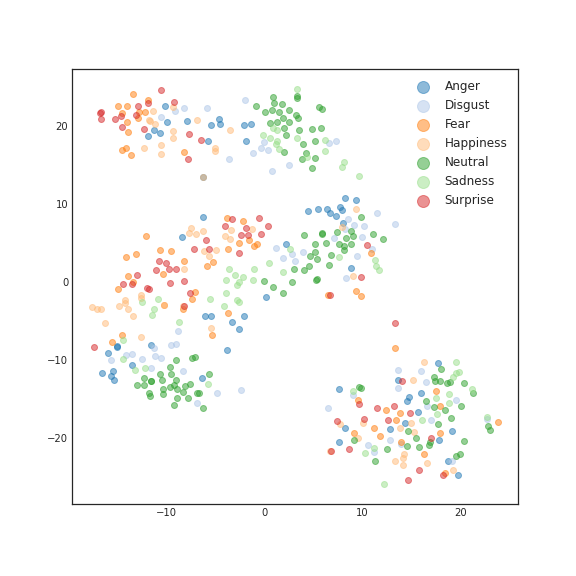}}\label{fig:ecapas}}
    \subfloat[Emo-DB]
    {{\includegraphics[width=0.25\linewidth,height =3.2cm ]{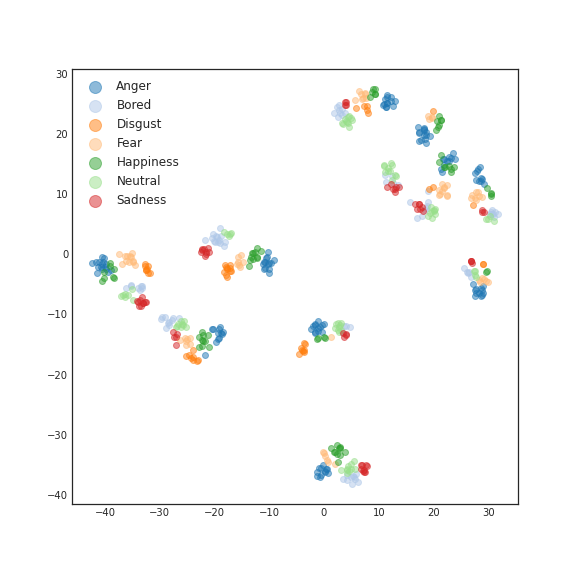}}\label{fig:ecapae}}
   \caption{t-SNE plots of ECAPA embeddings across different speech emotion corpora}
\label{fig:tsneecapa}
\end{figure*}

Embeddings derived from PTMs capture the semantic and aural information of the input clip. We intend to evaluate how effective these embeddings are at capturing emotional content by comparing embeddings retrieved from various PTMs. For selecting PTMs whose embeddings are to be used in our study, we follow two benchmarks: Speech processing Universal PERformance Benchmark (SUPERB) \cite{yang2021superb} and Holistic Evaluation of Audio Representations (HEAR) \cite{turian2022hear}.\par

SUPERB consists of various speech-related tasks ranging from speaker identification, speech emotion recognition, speech recognition, voice separation, speaker diarization, etc. We select models with top performance in SUPERB and are openly available such as wav2vec 2.0, data2vec, wavLM, and UniSpeech-SAT. For wav2vec 2.0, we choose the base\footnote{\url{https://huggingface.co/facebook/wav2vec2-base}} version for our experiments that contains 12 transformer blocks in its architecture. On SUPERB, data2vec delivers slightly lower results than the model with the best performance i.e wavLM. data2vec \cite{baevski2022data2vec} aims for bridging the gap in learning methods by proposing a generalized learning framework for different input modalities. wavLM \cite{chen2022wavlm} outperforms every other counterpart except UniSpeech-SAT on SUPERB. UniSpeech-SAT is a contrastive loss model with multitask learning. UniSpeech-SAT pre-training is done in a speaker-aware format whereas wavLM learns masked speech prediction and denoising concurrently during pre-training. This assists wavLM in dealing with multidimensional information contained in speech, such as speaker identity, spoken content, and so on. wavLM base+\footnote{\url{https://huggingface.co/docs/transformers/model_doc/wavlm}} version is used for carrying out our experiments and it is made of a total of 12 transformer encoder layers and was pre-trained on 94k hours data from various diverse speech databases including LibriLight, VoxPopuli, and GigaSpeech. We choose the base+ version for wavLM as it has achieved slight improvement over the base version on SUPERB with a similar number of parameters. For wav2vec 2.0, data2vec, wavLM, and UniSpeech-SAT the last hidden states are extracted and with the application of pooling average, they are converted to a vector of 768-dimension for each audio file to be used as input features for low-level classifiers. The input audio is sampled to 16KHz for all the self-supervised PTMs. We work with the base versions of wav2vec 2.0, data2vec\footnote{\url{https://huggingface.co/docs/transformers/model_doc/data2vec}}, and UniSpeech-SAT\footnote{\url{https://huggingface.co/docs/transformers/model_doc/unispeech-sat}} due to computational constraints and they were pre-trained on 960 hours of speech data from LS. wav2vec 2.0 is the lowest-performing model on SUPERB among all the self-supervised models under consideration, however, it has been applied for SER and proven to be effective in both English and multilingual formats \cite{sharma2022multi}. \par

As SUPERB is primarily concerned with speech processing tasks, PTMs pre-trained on speech data and in self-supervised manner, we chose various other PTMs with presence in HEAR such as wav2clip and YAMNet. Presence of wav2vec 2.0 can also be seen in HEAR leaderboard. wav2clip and YAMNet doesn't achieve SOTA performances on HEAR leaderboard and are mostly dominated by transformer-based architectures pre-trained in a self-supervised fashion. However, we added them in our evaluation as we wanted to access the effectiveness of their embeddings for SER as they were pre-trained using different methodologies and differed in terms of the data used for pre-training. wav2clip\footnote{\url{https://pypi.org/project/wav2clip/}} \cite{wu2022wav2clip} is pre-trained using knowledge distillation from CLIP and employs ResNet-18 as an audio encoder and uses VGGSound, an audio-visual Youtube database as pre-training data. Each audio file is transformed to a 2D sprectrogram for input to ResNet and converted to a vector of 512-dimension by average pooling. Similar to its parent architecture CLIP, wav2clip also transfers the audio embeddings to a joint embedding space. wav2clip embeddings as input features with supervised models have shown to be better than representations from other PTMs pre-trained on audio data in most datasets except FSD50K, where YAMNet representations performed better. YAMNet\footnote{\url{https://github.com/tensorflow/models/tree/master/research/audioset/yamnet}} pre-training is done in a supervised fashion on AS mainly for audio classification and is based on MobileNet V1 CNN architecture. YAMNet generates frame-level embeddings that are average pooled into 1024-dimension clip-level embeddings.\par

To broaden our assessment, we also considered PTMs for speaker recognition, as knowledge gained for speaker recognition can be beneficial for SER. Evidence suggests that information gained for speaker recognition can help in SER \cite{pappagari2020x}. Researchers have also advocated inserting knowledge about the speaker identity to network devoted to the primary job of SER \cite{moine2021speaker} to boost performance for SER. So, we select x-vector \cite{snyder2018x} and ECAPA \cite{desplanques2020ecapa} to validate the efficacy of speaker recognition system for SER. x-vector, a time delay neural network (TDNN) improves over previous speaker recognition system, i-vector and Emphasized Channel Attention, Propagation and Aggregation (ECAPA) approach inserts several modifications to the x-vector model architecture. We pick off-the-shelf x-vector\footnote{\url{https://huggingface.co/speechbrain/spkrec-xvect-voxceleb}} and ECAPA\footnote{\url{https://huggingface.co/speechbrain/spkrec-ecapa-voxceleb}} models. Both were pre-trained on a combination of voxceleb1 and voxceleb2 in a supervised manner. For pre-training of x-vector and ECAPA, all of the input audio files were sampled at 16Khz single-channel. We extract 512 and 192-dimension embeddings using Speechbrain \cite{speechbrain} for x-vector and ECAPA respectively.

\section{Experiments}

\label{exp}

\subsection{Downstream Classifier}

We experiment with two classical machine learning approaches XGBoost (XGB), and Random Forest (RF), and a fully connected network (FCN). FCN is a simple neural network with three dense layers, batch normalization and dropout in between. Activation function being used is \textit{relu} in all the dense layers and followed by \textit{softmax} in the output layer which outputs the probabilies for different emotions. The same models are trained and evaluated with all the embeddings taken under consideration. \par

All four speech emotion corpora are splitted to 85:15 ratio with 15\% being used for testing. Out of the remaining 85\%, 10\% is kept for validation and the rest is used for training the classifiers. Hyperparameters are selected based on the performance of the classifiers on the validation set using GridSearchCV from \textit{sklearn} library. 
We train the FCN for 50 epochs with a learning rate of 1e-3 and the optimizer being used is \textit{Adam}. In addition, learning rate decay and early stopping are also applied while training the FCN.

\subsection{Experimental Results}

We compared the performance of eight PTMs embeddings across four speech emotion databases with two popular metrics accuracy and F1-score (macro). Table \ref{tab:compxgb}, Table \ref{tab:comprf} and Table \ref{tab:compfcn} shows the results of XGB, RF, and FCN for different PTMs embeddings across different datasets respectively. \par

Among self-supervised embeddings (wav2vec 2.0, data2vec, wavLM, UniSpeech-SAT), UniSpeech-SAT performed the best. It achieved the highest performance on CREMA-D, TESS, Emo-DB in Table \ref{tab:compxgb} followed by CREMA-D, TESS, Emo-DB  in Table \ref{tab:comprf}, and CREMA-D, TESS, SAVEE, Emo-DB  in Table \ref{tab:compfcn}. Speaker-aware pre-training may have contributed to these findings. The second is wavLM embeddings that outperformed UniSpeech-SAT embeddings on SAVEE in Table \ref{tab:compxgb} and \ref{tab:comprf}. A diverse dataset and the approach for pre-training where denoising is concurrently involved might adhere to this outcome. Among data2vec and wav2vec 2.0, data2vec embeddings perform better than wav2vec 2.0, however, the data used for pre-training belongs to the same dataset (LS), this can be the result of the architectural difference between data2vec and wav2vec 2.0. \par

wav2clip embeddings perform better than the self-supervised embeddings excluding UniSpeech-SAT across almost all the databases. This could be resultant of the learned knowledge achieved from CLIP and also during its pre-training in a multi-modal format which aims to push all the modalities to a single embedding space. YAMNet embeddings achieved moreover comparable results w.r.t its self-supervised counterparts, sometimes higher and sometimes lower, for example, in Table \ref{tab:compxgb}, YAMNet embeddings proved to be more effective in capturing emotion on TESS and Emo-DB. YAMNet reported lower performance than wav2clip across all the datasets except only in one instance in Table \ref{tab:compfcn}. \par

Embeddings from speaker recognition PTMs outperformed all other embeddings from different speech/audio PTMs across all spoken emotion datasets. This might be a manifestation of the information learned to identify speakers, where it is trained to recognize and distinguish between unique speakers. As a result, they learned to recognize distinctive elements of a person's speech patterns, such as rhythm, tone, and pitch, as well as linguistic and behavioral variables. x-vector achieves the top performance in comparison to ECAPA in most instances except on TESS and Emo-DB in Table \ref{tab:comprf} and  \ref{tab:compfcn}. \par

We also present t-SNE plots of raw embeddings extracted from different PTMs to understand the emotion-wise cluster. Figures \ref{fig:tsnewav2vec2}, \ref{fig:tsnedata2vec}, \ref{fig:tsnewavLM}, \ref{fig:tsneunispeechsat}, \ref{fig:tsnewav2clip} \ref{fig:tsneyamnet}, \ref{fig:tsnexvector}, and \ref{fig:tsneecapa} illustrates the t-SNE plots for wav2vecv 2.0, data2vec, wavLM, UniSpeech-SAT, wav2clip, YAMNet, x-vector, and ECAPA embeddings respectively. These figures support the results obtained from the tables above, it can be seen the embeddings extracted from PTMs for speaker recognition have far better emotion clusters with the highest distance between them than embeddings from other PTMs, especially for TESS corpus followed by wav2clip, YAMNet, and UniSpeech-SAT embeddings. For CREMA-D and TESS, the clusters formed by all eight PTM embeddings are almost inseparable. The results from the tables as well as the t-SNE plots show that models pre-trained with knowledge of the speaker performs best in SER, as evidenced by the performance of UniSpeech-SAT among self-supervised PTMs and the overall performance of x-vector and ECAPA.

\section{Conclusion}

\label{conc}

PTMs have been useful in various speech and audio-related tasks. Pre-train it on vast amount of labeled or unlabeled data and these models or the derived features from it can be highly beneficial for a wide range of tasks. Out of the variety of speech processing tasks, SER is a hard task to recon with, as due to various factors comes into play including difference in voice, tone, and accent. Past literature have shown the usage of different speech/audio PTMs embeddings for SER. However, previous studies haven't presented an extensive comparison of PTMs for SER with inclusion of various perspectives such as architectures of the PTMs, data utilized during pre-training phase, and pre-training technique followed. Our studies tries to narrow down this research gap by comparing embeddings derived from eight PTMs (wav2vec 2.0, data2vec, wavLM, UniSpeech-SAT, wav2clip, YAMNet, x-vector, ECAPA) by training three classifiers (XGB, RF, FCN) on top of these features for four speech emotion datasets (CREMA-D, TESS, SAVEE, Emo-DB). Classifiers trained on embeddings extracted from models pre-trained for speaker recognition attained top performance in all corpora. Our findings suggest that the knowledge acquired for speaker recognition, such as recognition of tone and accent, provides benefits for SER. Embeddings generated from self-supervised PTMs have achieved SOTA performance across a wide range of downstream applications, with architectures such as wavLM and UniSpeech-SAT coming out on top. However, the results of our investigation show that embeddings from simpler CNN PTM like YAMNet still hold solid ground in terms of performance for SER. The outcomes of this study can be used to guide future studies in selecting appropriate embeddings for speech-emotion detection applications.

\noindent\textbf{Future Work:} We considered eight PTMs, and in the future, we plan to extend our work by incorporating more diverse speech/audio PTM architectures. We investigated four speech emotion corpora in this study, three in English and one in German; in the future, we aim to include more databases not just in English but also in other languages.

\bibliographystyle{IEEEtran}
\bibliography{references}
\label{sec:References}
\end{document}